\documentclass[
	aps, aps,pra,longbibliography, superscriptaddress, twocolumn,
	10pt
	floatfix, 
    nofootinbib,
	tightenlines
]{revtex4-1}
\usepackage[final]{graphicx}
\usepackage{times,bbm,amsmath,amssymb}
\usepackage{epsfig,color}
\usepackage{xcolor}
\usepackage{hyperref}
\hypersetup{
    colorlinks = true
}
\usepackage{cleveref}
\usepackage{microtype}

\usepackage{float,siunitx}
\usepackage[caption = false]{subfig}

\usepackage[greek,english]{babel}
\usepackage{thumbpdf,enumerate}
\usepackage{booktabs}
\usepackage{sidecap}
\usepackage[scaled=.8]{couriers}
\usepackage{multirow}
\usepackage{placeins}
\usepackage{relsize}
\usepackage{pst-grad,bm}
\usepackage{epigraph}
\usepackage{gensymb}
\usepackage{longtable}
\usepackage{ulem} 
\usepackage{acronym}
\usepackage{physics}
\usepackage{easyReview}

\usepackage[columnwise]{lineno}

\newcommand{\ale}[1]{{\color{black}#1}}

\DeclareUnicodeCharacter{0301}{\'{e}}

\begin{document}


\vspace{10pt}

\title{Quantum phase imaging via diffraction of correlated bi-photon states.}

\author{Nazanin Dehghan}
\address{Nexus for Quantum Technologies, University of Ottawa, Ottawa, K1N 6N5, ON, Canada}
\affiliation{National Research Council of Canada, 100 Sussex Drive, K1A 0R6, Ottawa, ON, Canada}

\author{Alessio D'Errico} 
\email{aderrico@uottawa.ca}
\address{Nexus for Quantum Technologies, University of Ottawa, Ottawa, K1N 6N5, ON, Canada}
\affiliation{National Research Council of Canada, 100 Sussex Drive, K1A 0R6, Ottawa, ON, Canada}

\author{Yingwen Zhang} 
\address{Nexus for Quantum Technologies, University of Ottawa, Ottawa, K1N 6N5, ON, Canada}
\affiliation{National Research Council of Canada, 100 Sussex Drive, K1A 0R6, Ottawa, ON, Canada}

\author{Benjamin Sussman} 
\affiliation{National Research Council of Canada, 100 Sussex Drive, K1A 0R6, Ottawa, ON, Canada}
\address{Nexus for Quantum Technologies, University of Ottawa, Ottawa, K1N 6N5, ON, Canada}

\author{Ebrahim Karimi}
\address{Nexus for Quantum Technologies, University of Ottawa, Ottawa, K1N 6N5, ON, Canada}
\affiliation{National Research Council of Canada, 100 Sussex Drive, K1A 0R6, Ottawa, ON, Canada}
\affiliation{Institute for Quantum Studies, Chapman University, Orange, California 92866, USA}

\begin{abstract}
Two-photon states generated through degenerate spontaneous parametric down-conversion (SPDC) can exhibit sharp correlations in the transverse spatial coordinates. This property leads to unique free-space propagation features. Here, we show that a phase object placed in the image plane of the source affects the free space propagation of the SPDC in a way that is mathematically analogous to the Fresnel diffraction of a first-order coherent source. This effect can be observed via the extraction of correlation images. We demonstrate this prediction with an experiment where the diffraction of correlated bi-photons is detected using an event-based camera. The results allow us to reconstruct the phase structure of the sample via non-interferometric phase retrieval methods. We verify that the retrieved phase patterns exhibit an enhanced contrast due to the probe’s two-photon nature. Our findings offer applications for non-interferometric, quantum-enhanced phase imaging.
\end{abstract}

\maketitle 

\section{Introduction}

Among the most intriguing applications of structured quantum light~\cite{walborn2010spatial,forbes2021structured,nape2023quantum,d2021quantum} is the possibility of exploiting correlations between optical modes in two or more photon states to boost the sensitivity of measurements beyond the limits allowed by classical states of light~\cite{barbieri2022optical}. Quantum metrology based on non-classical light has found applications exploiting squeezed states~\cite{bondurant1984squeezed, schnabel2017squeezed}, N$00$N states~\cite{mitchell2004super,xiang2011entanglement,hong2021quantum}, as well as induced coherence~\cite{mandel1995optical} to enhance phase sensitivity in optical interferometers or imaging apparatuses~\cite{hudelist2014quantum,frascella2019wide,thekkadath2020quantum,camphausen2021quantum, black2023quantum}, to increase the resolution of imaging systems~\cite{padgett2017introduction,moreau2018resolution, zhang2019interaction} and to resolve coherent~\cite{d2013photonic, barboza2022ultra} or incoherent~\cite{grenapin2022super} displacements. The most used probe for quantum imaging is the quantum state generated by spontaneous parametric down-conversion (SPDC). The non-classical features of this state allow accessing a broad range of quantum-enhanced measuring strategies using correlations in polarization~\cite{mitchell2004super,camphausen2021quantum, defienne2021polarization}, photon number~\cite{meda2017photon}, spatial modes~\cite{mair2001entanglement,simon2012two,hiekkamaki2021photonic,d2021full}, transverse position and/or momentum~\cite{chen2019realization,walborn2010spatial} as well as in frequency~\cite{chan2009two}.

Towards imaging applications, spatial correlations in position and anti-correlations in momentum spaces have been exploited in quantum ghost imaging experiments and can be extended to the two-color regime when spectral correlations are also employed~\cite{chan2009two}. While ghost imaging is based on triggering a pixelated detector with a bucket detector, other experiments make use of either post-selection on single spatial modes~\cite{fickler2013real} or of the reconstruction of spatially resolved coincidences. The latter approach has sparked a wide range of imaging applications based on coincidence analysis: sub-shot-noise quantum imaging \cite{brida2010experimental,lopaeva2013experimental,ortolano2023quantum}, quantum digital holography~\cite{camphausen2021quantum,defienne2021polarization,thekkadath2023intensity,zia2023interferometric}, spatially and spectrally resolved Hong-Ou-Mandel interferometry~\cite{ndagano2022quantum,zhang2021high,gao2022high}, pixel super-resolution imaging~\cite{toninelli2019resolution,defienne2022pixel}, light-field microscopy~\cite{dangelo2016correlation,zhang2022ray,Zhang2024}, hyperspectral imaging~\cite{zhang2023snapshot}, and high-dimensional quantum state characterization~\cite{zia2023interferometric,gao2024full,dehghan2024biphoton}. Most of these works used camera technologies, which effectively allow to detect spatially resolved coincidences without the need for raster scanning; these are based on EMCCD cameras~\cite{bolduc2017acquisition}, SPAD arrays~\cite{eckmann2020characterization} and event-based intensified cameras~\cite{nomerotski2023intensified, kundu2024high}. In particular, event-based cameras allow for a fast acquisition rate with the ability to resolve spatial correlations in fractions of seconds, thus allowing for data collection in relatively short times when compared with other detectors. 
Event-based cameras have been employed for developing new methods of quantum state characterization based on interferometry~\cite{zia2023interferometric}, spatially resolved polarization tomography~\cite{gao2024full} and non-interferometric bi-photon state phase retrieval~\cite{dehghan2024biphoton}. 
 
The latter approach showed that by post-selecting the distribution of two-photon coincidences on spatially correlated events, it is possible to isolate the contribution to the bi-photon state due to the pump amplitude from the contribution due to the phase matching function. This effect is related to the fact that the pump field coherence properties are transferred in the two-photon spatial correlations, as demonstrated experimentally in Refs.~\cite{verniere2024hiding, dehghan2024biphoton}. 
 
In this work, we demonstrate how the analysis of coincidences can be used to reveal the diffraction of correlated bi-photon states through a transparent sample with spatially varying thickness. The proposed technique is based on placing the sample in the near-field of the SPDC (nonlinear) crystal and collecting spatially resolved coincidence events at different planes away from the sample. 
 
Typically, observing the spatial distribution of two-photon coincidences, called the coincidence image, would not allow one to discern the phase structure of the sample. This is because the coincidence image is only a particular projection of the 4D coincidence distribution \cite{zia2023interferometric}. However, we show that by post-selecting the coincidences dataset on spatially correlated events, the resulting image is formally equivalent to the Fresnel diffraction pattern of a coherent light beam diffracted from the sample. Notably, the phase imparted by the sample appears doubled, meaning that the phase shift induced by the SPDC light with wavelength $\lambda$ matches that which would be imparted by the pump laser, with wavelength $\lambda/2$. This phase enhancement effect is due to the two-photon spatial correlations. 
The collected diffraction data can then be exploited to retrieve the phase pattern imprinted by the sample on the probe light. We show this by employing two different phase retrieval techniques -- the Gerchberg-Saxton (GS) algorithm and transport of intensity equation (TIE) -- applied to either an SPDC or a coherent probe, in order to demonstrate the phase contrast enhancement in the case of correlated bi-photons.
 
These results, integrated with quantitative phase retrieval techniques, could lead to novel applications in quantum-enhanced phase imaging.
\begin{figure}[t!]
\includegraphics[width=0.9\columnwidth]{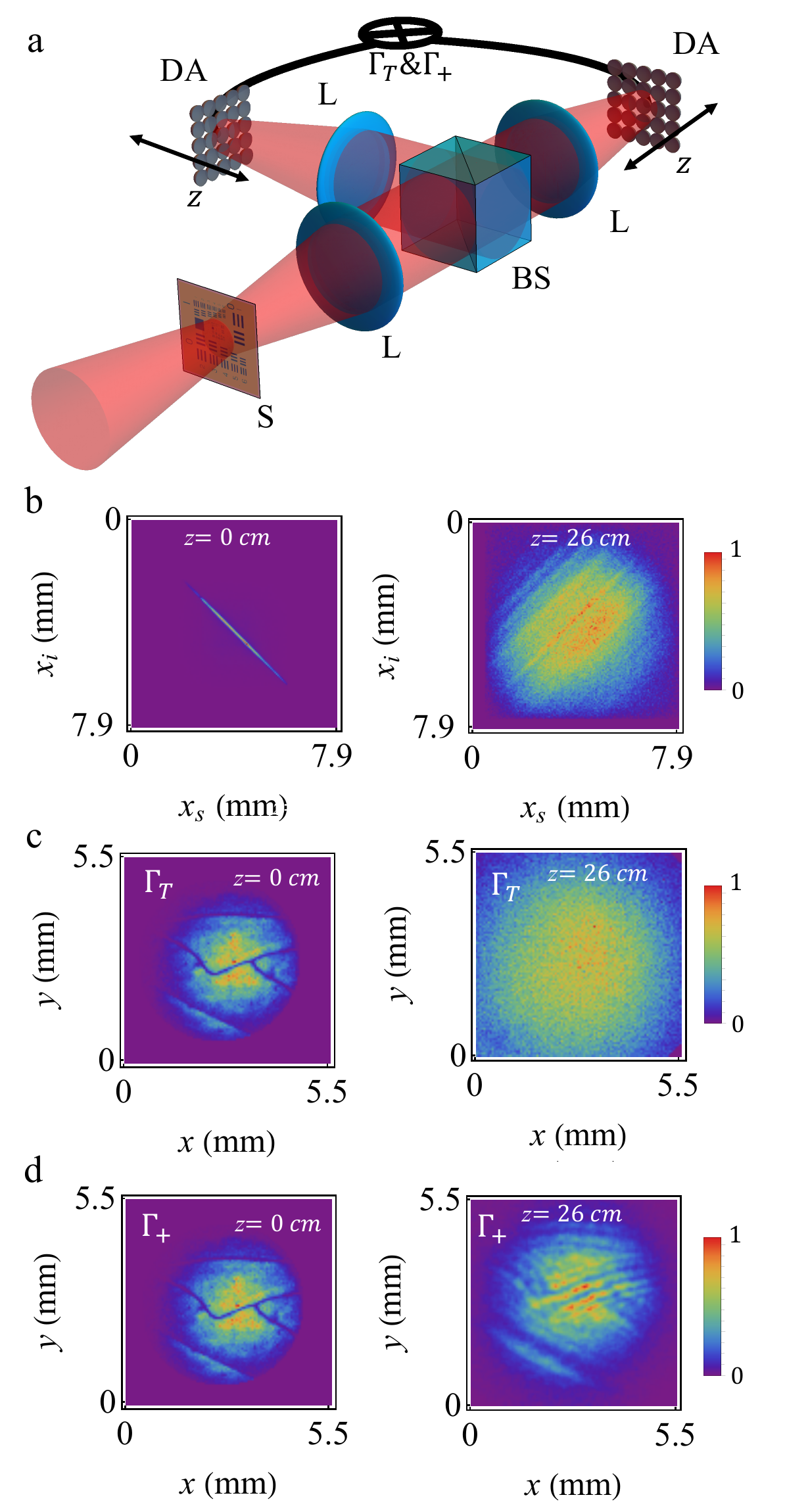}
\caption{\textbf{Extraction of correlation image.} \textbf{a}: Layout of a biphoton diffraction imaging experiment. S: sample. DA: detector array. The longitudinal position $z$ of the DAs can be varied in such a way as to allow for performing coincidence analysis in different propagation planes. BS: 50/50 beam splitter. L: lens. \textbf{b}: Experimentally observed $x$-correlations in a plane that is $0$\,cm and $26$\,cm away from the sample ($y$ direction correlations have a similar appearance). Sharp correlations are observed in the sample image plane. After $26$\,cm of free space propagation, the photon pair correlations are much broader due to the rapid divergence of the phase-matching function. \textbf{c}: Coincidence image, created from photon pairs that are detected within a $10$~ns coincidence time window. The sample structure is visible only in the near-field where position correlations are sufficiently narrow, and it completely disappears at larger propagation distances. \textbf{d}: Correlation image $\Gamma_+$, where signal and idler photons are post-selected at the same transverse position, up to a fixed shift. While in $z=0$ cm $\Gamma_+$ yields the same information as $\Gamma_T$, after free space propagation, as a consequence of Eq.~\eqref{eq:fresnelE}, $\Gamma_+$ exhibits a typical diffraction pattern, not visible in standard coincidence images. The imaged sample is a portion of a fly wing. }
\label{fig:postselection}
\end{figure}

\section{Theory}
Two-photon states with high spatial correlation can be generated by pumping a thin nonlinear crystal (thickness $\sim 1$ mm) with a collimated pump laser. In the degenerate case, where the two photons are emitted at the same frequency, the two-photon wavefunction in the transverse momentum space reads~\cite{walborn2010spatial, zia2023interferometric}:
\begin{align} \label{eq:wavefunction}
    \psi(\mathbf{k}_s,\mathbf{k}_i)=\text{sinc}\left(\frac{L\lambda_p}{4\pi}\abs{\mathbf{k}_s-\mathbf{k}_i}^2+\xi\right)\mathcal{A}(\mathbf{k}_s+\mathbf{k}_i),
\end{align}
where $L$ is the crystal thickness, $\lambda_p$ the central wavelength of the pump laser, $\mathbf{k}_{s,i}=(k_{x_{s,i}},k_{y_{s,i}})$ refers to the transverse wavevector components of signal ($s$) and idler ($i$) photons, $\xi$ is the longitudinal walk-off which can be tuned changing the orientation of the optical axis of the crystal, and $\mathcal{A}$ is proportional to the angular spectrum amplitude of the pump laser field. The sinc function corresponds to the phase-matching contribution to the bi-photon wavefunction. Equation~\eqref{eq:wavefunction} is equivalent to the far field structure of the bi-photon wavefunction, and neglects transverse walk-off effects, which is a good approximation for thin crystals and collimated pump beams. The spatial structure of the near-field can thus be derived by the Fourier transform of  $\psi(\mathbf{k}_s,\mathbf{k}_i)$. The phase matching in the near-field will give a narrow contribution of width $\sim \sqrt{L \lambda_p/4\pi}$, which is typically of the order of $\sim 30 \,\mu$m. 
As the event-based camera used in this experiment has a pixel size of 55 $\mu$m,  we can use a delta function to approximate the Fourier transform of the sinc function, and the bi-photon state in the near-field becomes
\begin{align}\label{eq:biphoton}
    \ket{\psi}=\int d^2x\,\mathcal{\tilde{A}}(\mathbf{x})\hat{a}_\mathbf{x}^{\dagger^2}\ket{0},
\end{align}
where $\hat{a}_\mathbf{x}^{\dagger}$ is a photon creation operator in the transverse position $\mathbf{x}$, $\mathcal{\tilde{A}}$ is the Fourier transform of $\mathcal{A}$, and $\ket{0}$ is the vacuum state. 
\ale{This state can be viewed as a continuous variable EPR state as shown in Ref. \cite{howell2004realization}. Here we highlight the fact that it can also be considered a multimode 2-photon N00N state. This is perhaps more evident if in Eq. \ref{eq:biphoton} we replace $\mathbf{x}\rightarrow(i,j)$, with $(i,j)$ labeling the sensor pixels ($i,j=1\ldots,N$), and assume $\tilde{\mathcal{A}}$ to be uniformly distributed. Then $\ket{\psi}=\frac{1}{N}\left(\ket{2_{1,1},0_{1,2},\ldots,0_{N,N}}+\ldots+\ket{0_{1,1},0_{1,2},\ldots,2_{N,N}}\right)=\sum_{i,j}\ket{0,\ldots2_{i,j},\ldots 0}/N$.  Such a highly correlated state is peculiar to SPDC states generated in free space, in contrast with attenuated coherent states that, even if considering only the two-photon contributions, would not exhibit the correlations as in Eq. \ref{eq:biphoton}. }

Consider a sample, with spatially variable transmittance and thickness (or variable refractive index), placed in the common paths of the signal and idler photons at the crystal near-field (see Figure \ref{fig:postselection}-a). The sample's action can be modeled as the following transformation on the creation operators:
\begin{align}    
    \hat{a}^\dagger_\mathbf{x}\rightarrow t(\mathbf{x})\hat{a}^\dagger_\mathbf{x}+i\sqrt{1-T(\mathbf{x})}\, \hat{a}^\dagger_\text{loss},
\end{align}
where $t(\mathbf{x}):=\sqrt{T(\mathbf{x})}e^{i\alpha(\mathbf{x})}$ is a function whose absolute value squared is the transmittance power $T(\mathbf{x})$ of the object and whose phase $\alpha(\mathbf{x})$ is related to the spatially variable thickness and/or refractive index. The term $\sqrt{1-T(\mathbf{x})}\, \hat{a}^\dagger_\text{loss}$ takes into account the reflected or absorbed photons, scattered in undetected modes, represented by the operator $\hat{a}^\dagger_\text{loss}$. The part of the transmitted state that contributes to coincidence measurements is,
\begin{align}\label{eq:sample_on_biphoton}
    \ket{\psi}=\int d^2x\, \mathcal{\tilde{A}}(\mathbf{x})t^2(\mathbf{x})\hat{a}_\mathbf{x}^{\dagger^2}\ket{0}.
\end{align}
The above equation shows how the effect of the sample on the bi-photon state is to modify the pump amplitude function $\mathcal{\tilde{A}}$ as $\mathcal{\tilde{A}}\rightarrow t^2 \mathcal{\tilde{A}} =T\mathcal{\tilde{A}} e^{i2\alpha}$, which appears as if the field passed through a sample with squared power transmittance and doubled phase. The squared power transmittance results in a contrast enhancement in the near-field coincidence images, a phenomenon observed, for instance, in Ref. \cite{wolley2022quantum}. The phase enhancement effect was first observed in a two-photon diffraction experiment in Ref. \cite{d2001two}. A coincidence image collected in the image plane would show only the effect of the power transmittance $T$ but not of the phase. To gain information about the phase, one could perform an interferometric measurement based on the technique presented in Ref.~\cite{zia2023interferometric} or observe the free-space propagation of the bi-photon state as in Ref.~\cite{dehghan2024biphoton}. Here, we will follow the second approach. As shown in Ref.~\cite{dehghan2024biphoton}, the paraxial propagation of the bi-photon wavefunction can be modelled as two independent Fresnel propagators applied, respectively, to the phase matching and the pump functions. More precisely, let $\mathbf{X}_{i,s}$ be the transverse coordinates in a plane at a distance $z$ from the crystal and $\mathbf{X'}_{i,s}$ the transverse coordinates in a plane at a distance $z'$. \ale{At arbitrary $z$, the bi-photon wavefunction has a form similar to Eq.~\eqref{eq:wavefunction}: $\psi_z(\mathbf{X}_{i},\mathbf{X}_{s})=\Phi_z(\mathbf{X}_-)\mathcal{E}_z(\mathbf{X}_+)$, with $\mathbf{X}_{\pm}:=\mathbf{X}_{i}\pm\mathbf{X}_{s}$. The bi-photon wavefunction in $z'$ can be obtained from the wavefunction in $z$ by applying a Fresnel propagator on $\Phi_z$ and $\mathcal{E}_z$:
\begin{align}
    \psi_{z'}(\mathbf{X'}_{i},\mathbf{X'}_{s})=&\Phi_{z'}(\mathbf{X}'_-)\mathcal{E}_{z'}(\mathbf{X}'_+)\cr
    \Phi_{z'}(\mathbf{X}'_-):=&\int e^{i \frac{\pi}{\lambda_p}\frac{|\mathbf{X}'_{-}-\mathbf{X}_-|^2}{(z'-z)}}\Phi_z(\mathbf{X_-})\,d^2X_-,\cr
    \mathcal{E}_{z'}(\mathbf{X_+'}):=&\int e^{i \frac{\pi}{\lambda_p}\frac{|\mathbf{X_+'}-\mathbf{X_+}|^2}{(z'-z)}}\mathcal{E}_z(\mathbf{X_+})\,d^2X_+
    \label{eq:biphotonsep}
\end{align}
%
%
%
In our case, $\mathcal{E}_{z=0}=t^2\mathcal{\tilde{A}}$, thus, the sample placed in the near-field will influence the propagation of the centroid contribution of the bi-photon, given, in general, by
\begin{align}\label{eq:fresnelE}
   \mathcal{E}_z(\mathbf{x'}):=\int e^{i \frac{\pi}{\lambda_p z}|\mathbf{x'}-\mathbf{x}|^2}t^2(\mathbf{x})\mathcal{\tilde{A}}(\mathbf{x})\,d^2x.
\end{align}}
\begin{figure*}
\includegraphics[width=1\textwidth]{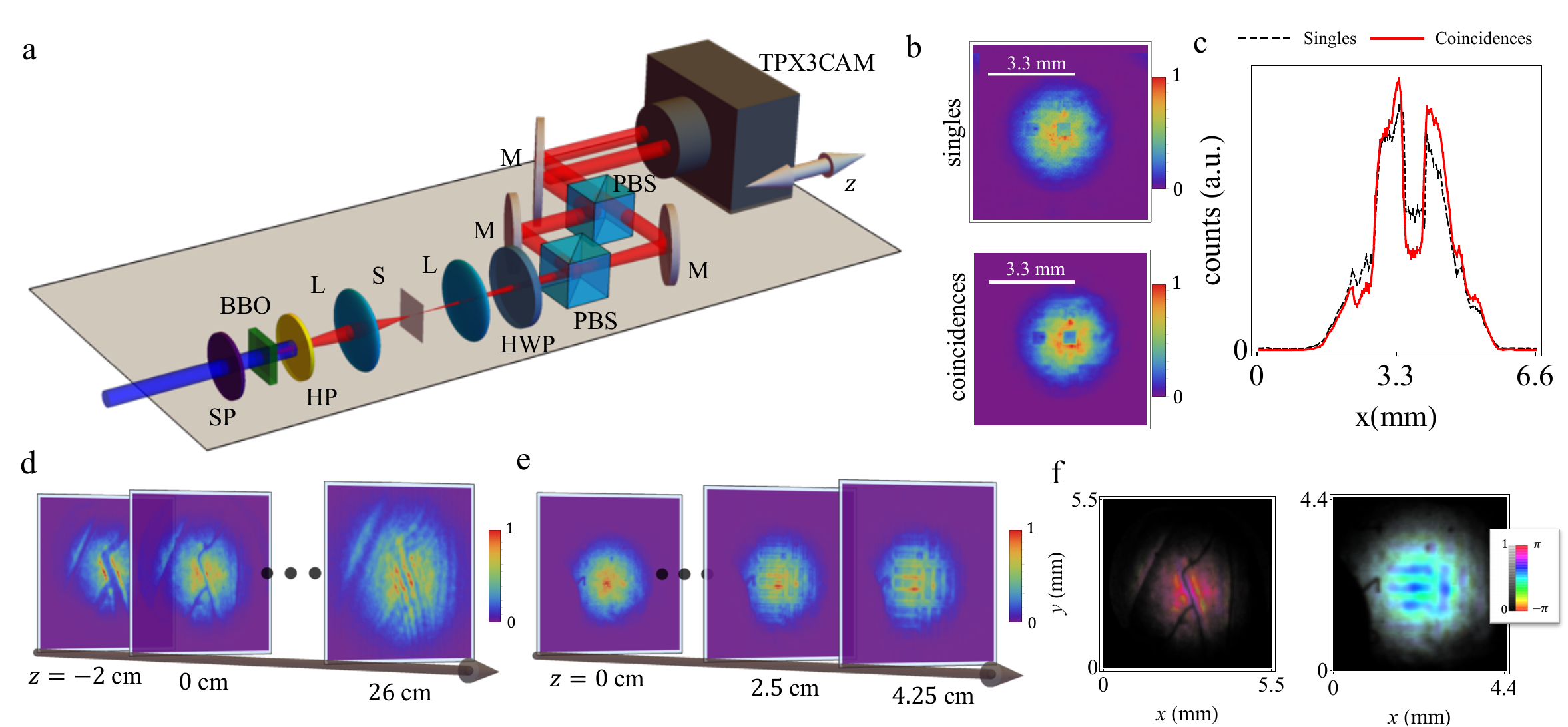}
\caption{\textbf{Experimental setup and extracted bi-photon diffraction patterns.} \textbf{a}: Sketch of the experimental setup. A collimated pump laser is incident on a Type-I BBO crystal and generates photon pairs. After removing the pump with a high-pass spectral filter (HP), the SPDC light is sent to a sample (S) placed in the image plane of the crystal. Using a half-wave plate (HWP) and polarizing beamsplitters (PBS), the idler and signal photons are split, with 50\% probability, to be detected on two different regions of an event camera (TPX3CAM). The camera is translated along the photon propagation direction, allowing for the acquisition of coincidence images at different propagation planes. M, Mirror. L, $4-f$ imaging represented as a single lens.
\textbf{b} and \textbf{c}: Contrast enhancement observed when comparing singles and coincidences images of the sample, a partially absorbing 350 nm thick square.
\textbf{d} and \textbf{e}: Diffraction patterns observed through correlation images of  
a portion of a fly wing \textbf{d} and  
for a 127\,nm thick resolution phase target \textbf{e}. 
\textbf{f}: Reconstructed phase and amplitude obtained by inversion of the TIE from the data collected at $z=0$ cm and $z=-2$ cm away from the wing sample (panel \textbf{d}), and at  $z=0$ cm and $4.25$ cm away from the resolution phase target (panel \textbf{e}) --similar results are obtained using the data at $z=2.5$ cm.}
\label{fig:setup}
\end{figure*}
The absolute value of this contribution can be experimentally isolated from $\Phi_z$ performing spatially resolved coincidence measurements and extracting the marginal distribution of coincidences with respect to the $\mathbf{X}_-$ variables. This procedure is detailed in Refs.~\cite{verniere2024hiding,dehghan2024biphoton} and briefly illustrated in Fig.~\ref{fig:postselection}: Coincidence events are recorded as a function of signal and idler coordinates \ale{$\mathcal{C}(\mathbf{X}_+,\mathbf{X}_-)=\abs{\Phi_z(\mathbf{X}_-)}^2\abs{\mathcal{E}_z(\mathbf{X}_+)}^2$}. The spatial distribution of events, called ``coincidence image'', will generally appear as in Fig.~\ref{fig:postselection}-c. The coincidence image corresponds to the marginal distribution $\Gamma_T(\mathbf{X}_s)=\int\mathcal{C}(\mathbf{X}_+,\mathbf{X}_-)d^2X_i$. In generic propagation planes, where the spatial correlations are broadened, $\Gamma_T(\mathbf{X}_s)$ does not clearly display the features of the sample due to the convolution with the phase matching function $\Phi$. However, one can integrate along the $\mathbf{X}_-$ coordinates, obtaining what, according to the nomenclature in Ref. \cite{verniere2024hiding}, is denoted as ``correlation image" (Fig.~\ref{fig:postselection}-d):
\begin{equation}
\Gamma_+(\mathbf{X}_+):=\int\mathcal{C}(\mathbf{X}_+,\mathbf{X}_-)d^2X_-=\abs{\mathcal{E}_z(\mathbf{X}_+)}^2.
\end{equation}
To extract the correlation image $\Gamma_+$ we use a slightly modified method, detailed in Ref. \cite{dehghan2024biphoton}, and in Supplementary Materials S1. Our method is based on creating separately a correlation image obtained by selecting only spatially correlated events where $\mathbf{X}_i=\mathbf{X}_s + \mathbf{c}$, with $\mathbf{c}$ a constant, and then summing them up, with a corresponding displacement on each sub-image to account for the shift $\mathbf{c}$. This method allows us to confirm the assumption that the biphoton wavefunction can be separated as in Eq. \ref{eq:wavefunction}. If Eq. \ref{eq:wavefunction} is verified, each sub-image will have the same pattern and differ from others only in terms of the maximum number of counts and local shot noise fluctuations. On the contrary, if Eq. \eqref{eq:wavefunction} is not verified -- for instance, in the presence of walk-off effects in the SPDC source--, each sub-image will exhibit a slightly different pattern, which results in a blurred correlation image $\Gamma_+$.  
\newline
Our results show that if the correlation image is extracted from data collected at a distance $z$ from the sample plane, $\Gamma_+$ will be given by the Fresnel diffraction pattern of the sample (Fig.\,\ref{fig:postselection}-d). It must be noted that the Fresnel propagation must be calculated assuming the wavelength of the pump laser in the propagator rather than the wavelength of the down-converted photon pair.
An intriguing consequence of this result is the fact that the structure of a phase-only object, while it would not be discernible from coincidence images at $z=0$, can be revealed by correlation images collected at a distance $z\neq0$. 
\begin{figure*}
\includegraphics[width=\textwidth]{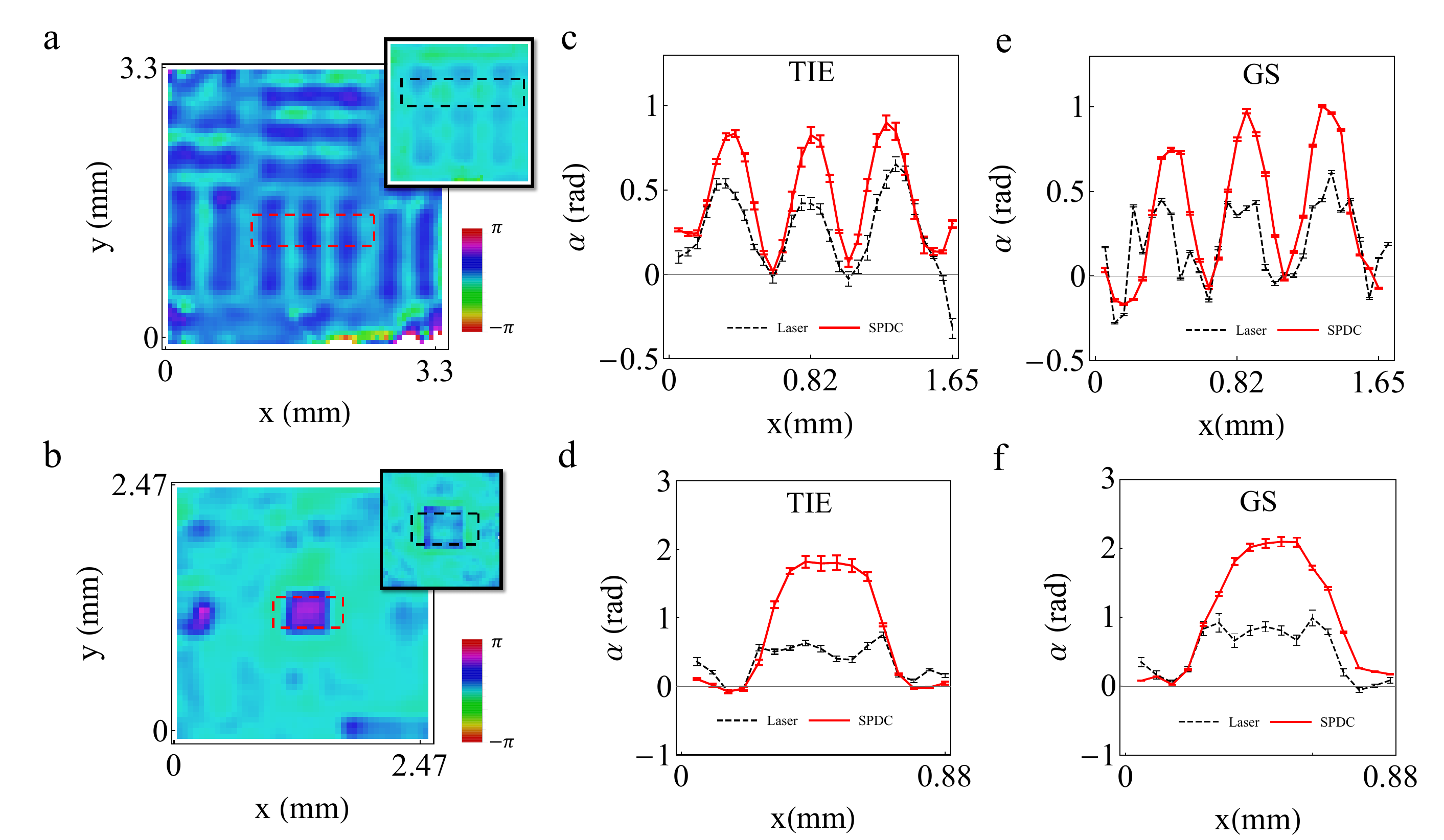}
\caption{
\textbf{Phase-enhancement from using the spatial correlations of SPDC.} \textbf{a}: Reconstructed phase pattern for a bi-photon from the image plane of a phase target with a depth of 127 nm (phase = 0.46). The inset shows the phase reconstructed using an 810 nm laser beam as a probe. \textbf{b}: Phase reconstruction by TIE for a bi-photon passing through a 350 nm deep square. 
The inset shows the reconstructed phase using a laser as a reference. All the reconstructions in \textbf{a} and \textbf{b} were obtained through inverting the TIE -- reconstructions using GS are shown in the Supplementary Information. The phase pattern here displayed for SPDC (laser) probe was obtained from correlation (intensity) images reported in Supplementary Figure S3.  \textbf{c} and \textbf{d}: Averaged cross-section along $y$ of the reconstructed phase through TIE within the dashed squares in \textbf{a} and \textbf{b} and provides a quantitative comparison between the phase obtained from SPDC and an 810\,nm laser. \textbf{e} and \textbf{f}: Comparison of the phases reconstructed through the GS algorithm. Error bars are standard deviations of the phase values obtained along the $y$ direction within the highlighted regions of interest. }
\label{fig:Uandpie}
\end{figure*}

\section{Experiment}
We verify our prediction on the bi-photon diffraction by a transparent object using an event camera (TPX3CAM). The TPX3CAM provides nanosecond temporal resolution on each of the $256\times 256$ pixels, thus, allowing the time and spatial correlations between SPDC photons to be measured simultaneously. The setup is sketched in Fig.~\ref{fig:setup}-a. A pump laser (405\,nm wavelength) is incident on a 1-mm-thick Type-I BBO crystal and generates frequency-degenerate (810\,nm) photon pairs via SPDC, spectrally filtered with a 10\,nm bandpass filter.
The photons are then used to illuminate a sample placed in the image plane of the crystal through a 4-f lens system. As a comparison to the bi-photon results, we have also performed the same imaging procedure classically by switching the pump laser with a 810\,nm laser.

Figure~\ref{fig:setup}-b shows the contrast enhancement observed in the image plane of the sample. A partially absorbing phase sample in the form of a 350\,nm thick square (Benchmark Technologies) is imaged, and a clear contrast enhancement is visible when comparing the singles and coincidence images. This is more quantitatively highlighted in the plot in panel c, where showing a cross-section of the square. 

Figure~\ref{fig:setup}-d and e show two examples of the correlation images $\Gamma_+$, reconstructed by translating the camera along the propagation direction $z$.  
Figure~\ref{fig:setup}-b shows the case of a fly wing as an example where the transmission function $t$ has a variable amplitude and phase, while  Fig.~\ref{fig:setup}-c shows the results for a phase-only sample, a phase calibration resolution target introduced in Ref.~\cite{Godden:16} with depth variations of $d=127$\,nm and refractive index $\sim 1.47$. For each plane, data was collected for 400 seconds. 
It is evident from Fig.~\ref{fig:setup}-d-e that the diffraction pattern due to the phase and amplitude variations becomes visible in subsequent propagation planes. We stress that these patterns can be observed only from the correlation images. No coherent-like propagation effects would be observed from first-order correlations or without appropriate analysis of the spatial correlations. Interestingly, our technique allows the retrieval of the structure of phase-only samples from the data collected in the near-field plane without performing interferometric experiments. Figure~\ref{fig:setup}-f shows the reconstructed near-field amplitude and phase using the correlation images collected at subsequent planes and employing phase retrieval techniques such as the Transport of Intensity Equation (TIE) (with results shown in Fig.~\ref{fig:setup}-f) or the Grechberg-Saxton (GS) algorithm (with results shown in Supplementary Figures S1 and S2), which we discuss in the following. 

Phase reconstruction can be performed using two approaches. The first was based on a modified version of the GS algorithm \cite{gerchberg1972practical}. While in the standard GS, one considers input data connected by a Fourier transformation, here we use a modified version of the GS where the Fourier transform is replaced with a Fresnel propagator (see Refs. \cite{Zhang2024, dehghan2024biphoton}). 
The GS algorithm can suffer from slow convergence and outputs corresponding to local minima. Moreover, it is better suited when the input data correspond to distant --ideally conjugate-- planes. However, in our context, it is difficult to reach this regime since the transverse extent of the SPDC signal is dominated by the phase matching function, leading to an inevitable loss of the resolution of the correlation images, if the data are collected far away from the crystal plane. A more suitable approach to our experimental scenario makes use of the transport of intensity equation (TIE) \cite{zuo2020transport,ortolano2023quantum} 
\begin{equation}
     \frac{dI}{dz}=\frac{\lambda}{2\pi} I \nabla_{\perp}^2 \phi,
 \end{equation}
where $\phi$ is the unknown phase, and $\nabla_{\perp}^2$ is the Laplacian in the transverse coordinates. The TIE is valid for paraxial propagation of fields with irrotational phase and uniform intensity and is equivalent to a Poisson equation for the phase with the source given by the longitudinal intensity variation. It can be thus inverted via Fourier transform and using the Green function of the Poisson equation,  $\mathcal{G}(\mathbf{k})=-2\pi /(\lambda \mathbf{k}^2)$, where $\mathbf{k}=(k_x,k_y)$ is the position vector in Fourier space. To remove the singularity at the origin of the spatial frequency space, we use the regularized kernel $\mathcal{G}_r:=-\lambda \mathbf{k}^2/[(\lambda\mathbf{k}^2)^2/2\pi+2\pi\epsilon]$, where $\epsilon$ is a regularization constant (note that $\mathcal{G}_r\rightarrow \mathcal{G}$ for $\epsilon\rightarrow0$). The phase was obtained from $\phi\approx \mathcal{F}^{-1}[\mathcal{G}_r \mathcal{F}[\Delta I/\Delta z]]/I_0$, where $\mathcal{F}$ and $\mathcal{F}^{-1}$ are the fast Fourier transform and its inverse, respectively, and $I_0$ the intensity measured in the sample plane. Being based on approximating $dI/dz$ with the experimentally collected increment $\Delta I/\Delta z$, the TIE is more suitable in our experiment, where we can easily observe diffraction between nearby planes. 
We chose values of the order of $\epsilon \sim20-40 \,(70-200)\, \text{cm}^{-2}$ for the biphoton (classical) scenario. The value of $\epsilon$ was chosen as the one that minimizes the mismatch between the experimental diffracted intensity $I_{exp}(z)$ ($z$ being the distance from the sample image plane) and the one obtained via a numerical propagation by a distance $z$ of the reconstructed field in the sample plane. In Supplementary Figure S4, we show how the reconstructed phase varies with the choice of $\epsilon$ with the corresponding error. 

We summarize the results of the TIE and GS approaches in Fig.~\ref{fig:Uandpie}.
Figures~\ref{fig:Uandpie} a-b show the reconstructed phase patterns using the TIE for two samples: the USAF resolution target with 127 nm phase variation and the square-shaped phase sample (Benchmark Technologies) of nominally 350 nm height. Qualitatively similar results are obtained by use of the GS algorithm (see Supplementary Figure S2 for the full phase patterns), showing how both approaches are useful to retrieve the phase image in the sample plane. 
Diffraction patterns used as input for the phase retrieval are shown in Supplementary Figure S3. 

Figure~\ref{fig:Uandpie} shows the recovered phase from imaging a pure phase sample with correlated bi-photons and is compared to that taken with a 810\,nm laser. Figure~\ref{fig:Uandpie}-c-d (using TIE) and Fig.~\ref{fig:Uandpie}-e-f (using GS) are the average phase in the regions highlighted in Fig.~\ref{fig:Uandpie}-a-b. The results show a phase enhancement of a factor of 2, which is in agreement with the predictions. The results in Fig.~\ref{fig:Uandpie} show that the phase pattern retrieved from SPDC diffraction appears to have less overall fluctuation compared to that retrieved using a classical laser, albeit having larger errorbars. We believe this is due to collecting the classical data with an 8-bit depth CMOS camera which does not having enough sensitivity to observe finer intensity fluctuations, thus resulting in a smaller uncertainty in the measured values. 

\section{Conclusions}
We have demonstrated that when SPDC states with sharp spatial correlations are transmitted through a transparent phase sample; the bi-photon state undergoes a second-order coherent diffraction effect. This effect can be mathematically treated as a Fresnel propagator applied to the pump field multiplied by the sample's transmittance squared. Phase information can thus be observed by collecting defocused coincidence images, provided that a postselection on spatially correlated photon pairs is performed.
A quantitative reconstruction of the phase structure can be obtained using phase retrieval methods. We used two methods (GS algorithm and Transport of Intensity equation) to reconstruct the sample's phase and demonstrate phase enhancement by a factor of 2 when comparing the bi-photon state with a laser beam of the same wavelength. We note that this phase enhancement effect can be used to improve the SNR of phase imaging techniques. In a similar work, Ref.~\cite{ortolano2023quantum}, sub-shot noise phase imaging with TIE was demonstrated, but with only one photon of the pair probing the sample, thus no phase and amplitude enhancement was observed. A quantum advantage in SNR could be achieved with our technique by using time-stamping cameras with higher detection efficiency (e.g. imaging devices based on SPAD arrays or superconductive nanowire detector arrays). If high detection efficiency is possible, it will be interesting in future studies to investigate the trade-off between exposure time and improved SNR to establish the range of photon flux where the use of quantum light can offer advantages in reducing the amount of noise affecting the performance of phase retrieval algorithms and, at the same time, keeping the sample at low exposure levels. Moreover, we expect that the noise tolerance of the phase retrieval algorithms considered here could be improved by integration with Neural Network architectures.

\section{Acknowledgments}
This work was supported by the Canada Research Chair (CRC) Program, NRC-uOttawa Joint Centre for Extreme Quantum Photonics (JCEP) via the Quantum Sensors Challenge Program at the National Research Council of Canada, and Quantum Enhanced Sensing and Imaging (QuEnSI) Alliance Consortia Quantum grant.

\bibliography{bibliography.bib}

\vspace{1 EM}

\noindent\textbf{Author Contributions}
N.D. and A.D. conceived the idea.  N.D., under the supervision of A.D and Y.Z., performed the experiment. Y.Z. prepared the code for the raw data analysis and A.D. for the correlation images extraction. N.D. and A.D. analyzed the data. B.S. and  E.K. supervised the project. A.D. prepared the first version of the manuscript. All authors contributed to the writing of the manuscript.
\vspace{1 EM}

\noindent\textbf{Data availability}
\noindent
The data that support the findings of this study are available from the corresponding author upon reasonable request.
\vspace{1 EM}

\noindent\textbf{Code avialability}
\noindent
The code used for the data analysis is available from the corresponding author upon reasonable request.

\vspace{1 EM}
\noindent\textbf{Ethics declarations} Competing Interests. The authors declare no competing interests.

\vspace{1 EM}
\noindent\textbf{Corresponding authors}
Correspondence and requests for materials should be addressed to aderrico@uottawa.ca.

\clearpage
\onecolumngrid
\begin{center}
{\Large Supplementary Material for: \\ Quantum phase imaging via diffraction of correlated bi-photon states.}
\end{center}
\vspace{1 EM}

\subsection*{Data analysis}

The analysis used to extract the pump and phase-matching contributions from the 4D coincidence distribution measured by the Timepix camera closely follows the approach in Ref. \cite{dehghan2024biphoton}. By performing coincidence detection, we identify photon pairs arriving at the camera simultaneously (within a 10 ns window). Considering only two-photon coincidences, we record the positions of the idler and signal photons.

This information enables various analyses of their spatial distributions. For instance, one can examine the marginal distributions: $\Gamma_y(y_i,y_s)=\iint{ dx_i dx_s \mathcal{C}(\textbf{X}_i,\textbf{X}_s)}$ or $\Gamma_x=\iint{ dy_i dy_s \mathcal{C}(\textbf{X}_i,\textbf{X}_s)}$. Another set of marginal distributions of interest are $\Gamma_i=\iint{ dx_s dy_s \mathcal{C}(\textbf{X}_i,\textbf{X}_s)}$ and $\Gamma_s=\iint{ dx_i dy_i \mathcal{C}(\textbf{X}_i,\textbf{X}_s)}$ which we refer to as coincidence images for idler and signal photon, respectively.

In the thin-crystal approximation for SPDC, the coincidence distribution can be expressed as $\mathcal{C}(\textbf{X}_i,\textbf{X}_s)=|\Phi(\textbf{X}_i-\textbf{X}_s)|^2 |\mathcal{E}(\textbf{X}_i+\textbf{X}_s)|^2$. By introducing the new variables $\textbf{X}_+ = \frac{\textbf{X}_i+\textbf{X}_s}{2}$, and $\textbf{X}_- = \frac{\textbf{X}_i-\textbf{X}_s}{2}$, we can rewrite the distribution as $\mathcal{C}(\textbf{X}_+,\textbf{X}_-)=|\Phi(\textbf{X}_-)|^2 |\mathcal{E}(\textbf{X}_+)|^2$. This change of variables naturally leads to another set of marginal distributions: $\Gamma_+(x_+,y_+)=\iint{ dx_- dy_- \mathcal{C}(\textbf{X}_+,\textbf{X}_-)}$, and $\Gamma_-=\iint{ dx_+ dy_+ \mathcal{C}(\textbf{X}_+,\textbf{X}_-)}$. Due to the factorized form of $\mathcal{C}(\textbf{X}_+,\textbf{X}_-)$ in this approximation, we obtain $\Gamma_+(x_+,y_+)=|\mathcal{E}(\textbf{X}_+)|^2$, and $\Gamma_-(x_-,y_-)=|\Phi(\textbf{X}_-)|^2$.

The acquisition of $\Gamma_+$ and $\Gamma_-$, and thus the separation of pump and phase-matching contributions, relies on the assumption that the bi-photon wave function factorizes into pump and phase-matching terms. In our analysis, we examine photons where $\mathbf{X_i} = \mathbf{X_s} + \mathbf{c}$, varying 
$\mathbf{c}$ to scan correlations and generate a set of images. If the wave function assumption holds, these images should be identical up to a shift determined by $\mathbf{c}$, differing only in photon count. By compensating for the shift and superimposing them, we enhance the statistics of the coincidence image and obtain the marginal distributions. However, any deviation from the assumption introduces variations among the images, causing blurring when they are overlaid. 

\subsection*{Supplementary Figures}

\begin{figure*}
\includegraphics[width=0.7\textwidth]{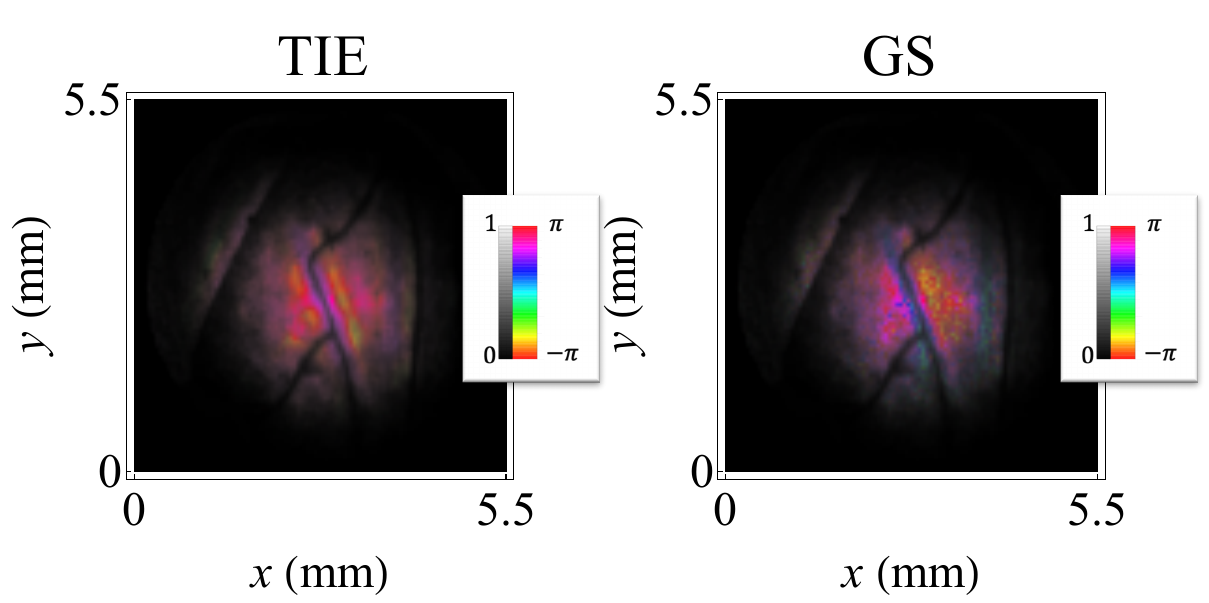}
\caption{\textbf{Reconstructed wing phase and amplitude patterns.} Complex transmittance of the wing sample obtained using the TIE or the GS algorithm, respectively.}
\label{fig:wingphase}
\end{figure*}

Figure \ref{fig:wingphase}
shows the reconstructed complex transmittance of the wing phase comparing the results obtained with TIE inversion and GS-algorithm, respectively.

\begin{figure*}
\includegraphics[width=0.7\textwidth]{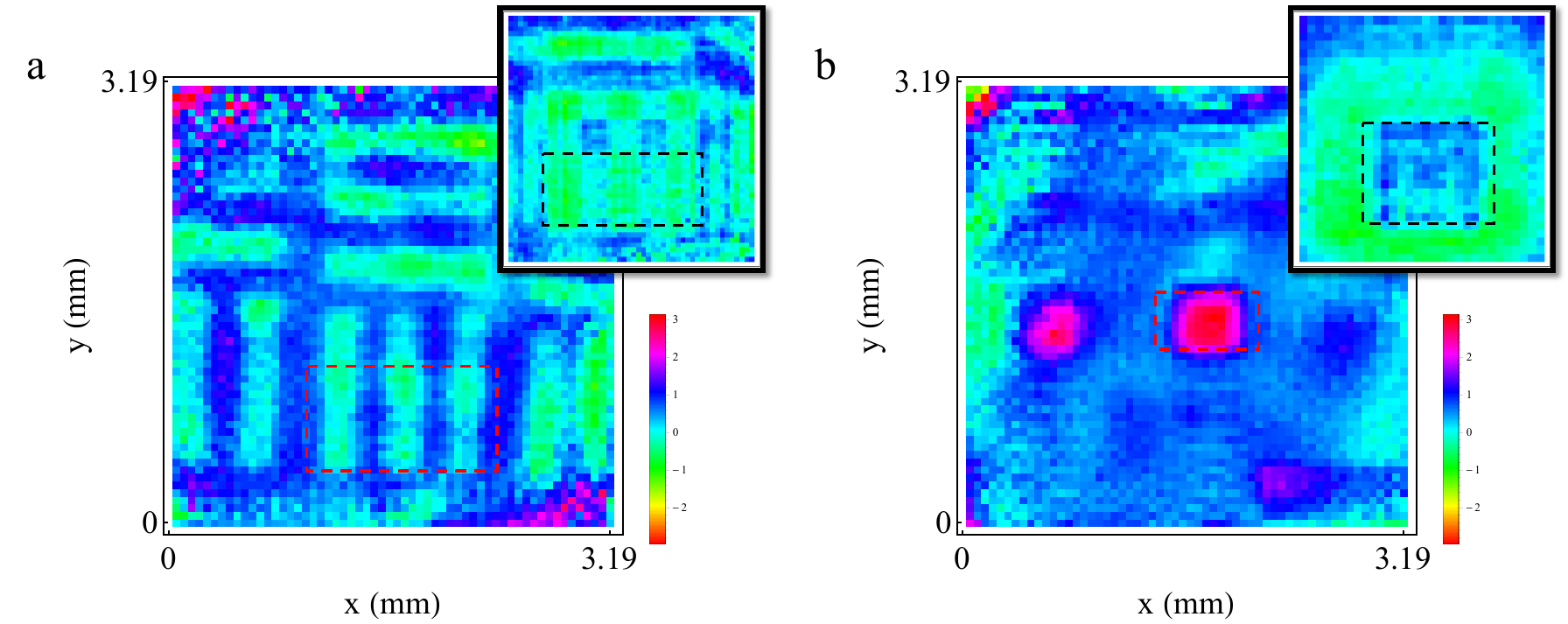}
\caption{\textbf{Reconstructed phase patterns from the GS algorithm.} \textbf{a}. Phase of the 127 nm target in the image plane using a bi-photon, with the inset showing the laser-based reconstruction. \textbf{b}. The same reconstruction for a 350 nm depth sample.}
\label{fig:GSResults}
\end{figure*}

Figure \ref{fig:GSResults}
shows the full field phase pattern obtained using the GS algorithm on data obtained using SPDC as a probe, as well as classical laser light (insets).

\begin{figure*}
\includegraphics[width=0.7\textwidth]{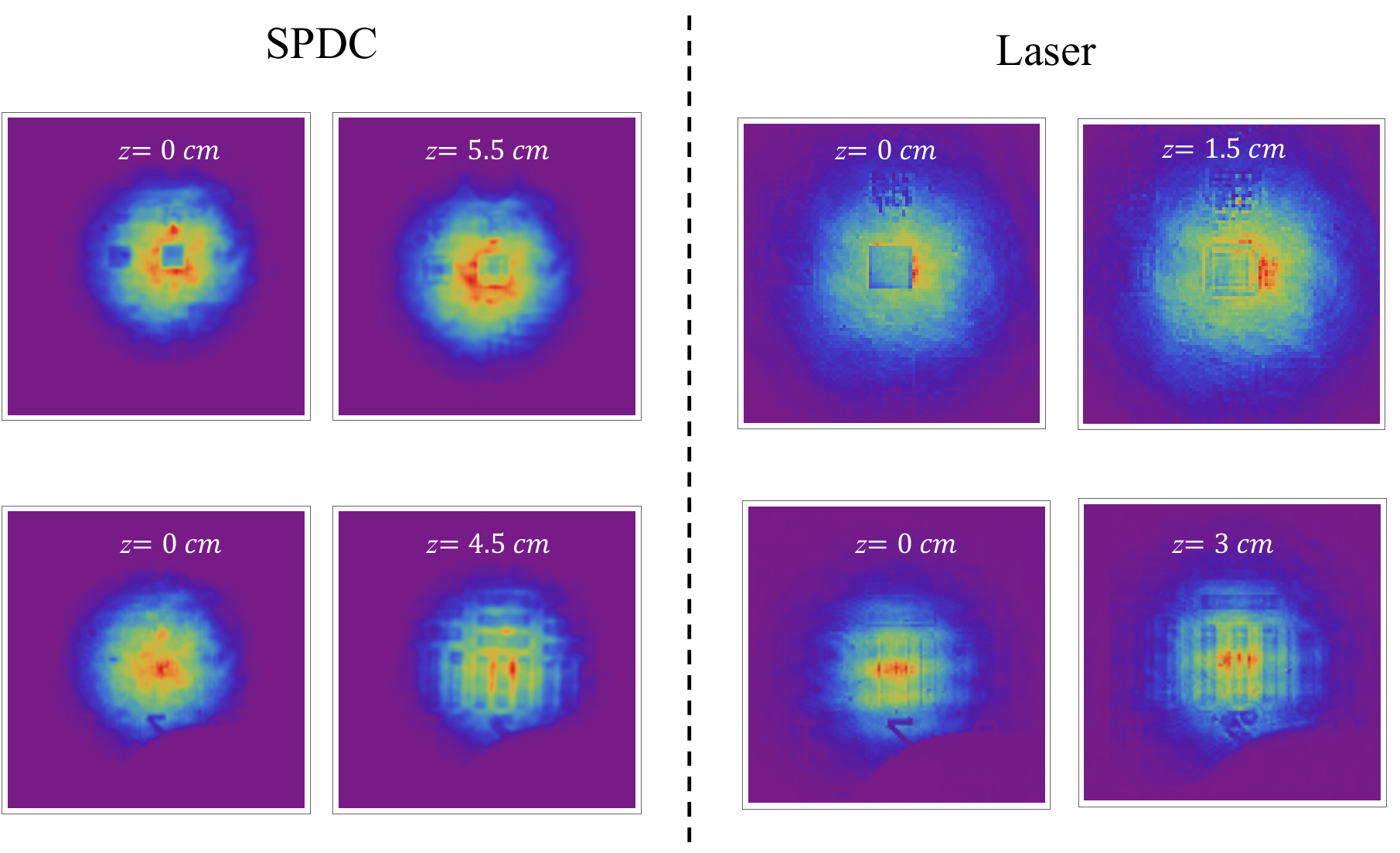}
\caption{\textbf{Input images used for phase retrieval with TIE.} We show the images used to obtain the phase pattern reported in the main text via inversion of the transport of intensity equation in both cases with a classical laser and SPDC probe. We used a value of $\Delta z$ larger in the SPDC to account for the fact that diffraction is dictated by the pump wavelength $\lambda_p$. }
\label{fig:tie_inputs}
\end{figure*}

Figure \ref{fig:tie_inputs} reports the inputs used to extract the phase with TIE reported in the manuscript.

\begin{figure*}
\includegraphics[width=\textwidth]{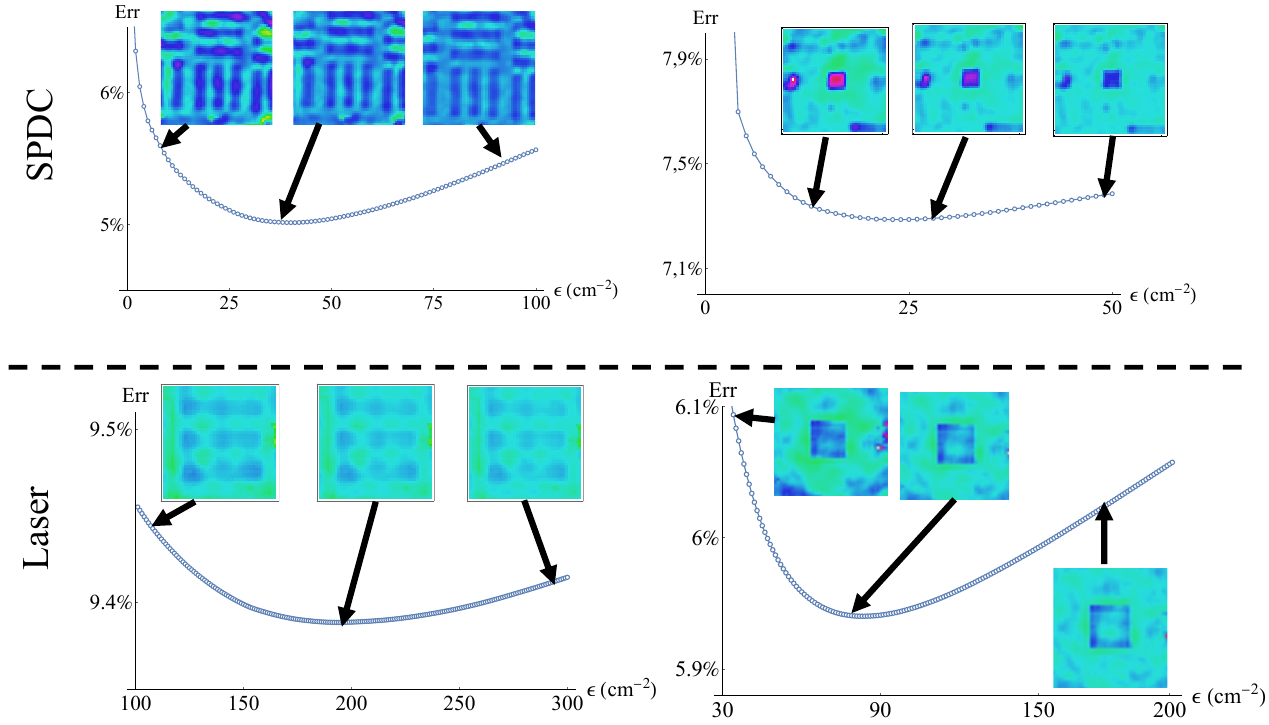}
\caption{\textbf{Effect of regularization constant $\epsilon$ on the reconsturcted phases.}}
\label{fig:tie_epsilon}
\end{figure*}

Figure  \ref{fig:tie_epsilon} shows the effect of the regularization factor $\epsilon$ in the kernel $\mathcal{G}_r$ on the retrieved phases with the TIE approach. Each panel shows the absolute error defined as $Err=\abs{I_{exp}(z_2)-I_{TIE}(z_2)}$, where $I_{exp}(z_2)$ is the experimental intensity in the second propagation plane, and $I_{TIE}(z_2)$ is given by:
\begin{align}
I_{TIE}:=\abs{\iint \mathbf{F}(z_2,z_1)(x',y')\sqrt{I_{exp}(z_1,x',y')}\exp{(i\phi_{TIE}(x',y'))}dx'dy' }^2.
\end{align}
Thus $I_{TIE}$ is the propagated intensity obtained assuming the phase obtained from the solution of the TIE as the field phase in the plane $z=z_1$. The optimal value of $\epsilon$ is the one that gives the minimum error $Err$.

\newpage

\newpage

\end{document}